\documentclass[twocolumn,english,aps,prl,showpacs,superscriptaddress]{revtex4-1}
\usepackage[T1]{fontenc}
\usepackage[latin9]{inputenc}
\usepackage{geometry}
\usepackage{color}
\usepackage{babel}
\usepackage{amsmath}
\usepackage{amssymb}
\usepackage{graphicx}
\usepackage[unicode=true,
 bookmarks=true,bookmarksnumbered=true,bookmarksopen=true,bookmarksopenlevel=1,
 breaklinks=false,pdfborder={0 0 0},pdfborderstyle={},backref=false,colorlinks=false]
 {hyperref}
\hypersetup{pdftitle={Your Title},
 pdfauthor={Your Name},
 pdfpagelayout=OneColumn,pdfnewwindow=true,pdfstartview=XYZ,plainpages=false}
\usepackage{breakurl}

\makeatletter
\usepackage{babel}

\usepackage{hyperref}
\def\equationautorefname~#1\null{Equation (#1)\null}

\makeatother

\begin{document}

\title{Light storage for one second at room temperature}

\author{Or Katz}
\email[Corresponding author:]{ or.katz@weizmann.ac.il }

\affiliation{Department of Physics of Complex Systems, Weizmann Institute of Science,
Rehovot 76100, Israel}

\affiliation{Rafael Ltd, IL-31021 Haifa, Israel}

\author{Ofer Firstenberg}

\affiliation{Department of Physics of Complex Systems, Weizmann Institute of Science,
Rehovot 76100, Israel}
\begin{abstract}
\textbf{Light storage, the controlled and reversible mapping of photons
onto long-lived states of matter \cite{Lukin-RMP-2003}, enables memory
capability in optical quantum networks \cite{oPTICAL-QUANTUM-COMPUTING-SCIENCE,Heshami-2016,Tittel-review-nature-photonics,Gisin-repeaters,Firstenberg-FLAME}.
Prominent storage media are warm alkali gases due to their strong
optical coupling and long-lived spin states \cite{Polzik-RMP-2010,Buchler-GEM}.
In a dense gas, the random atomic collisions dominate the lifetime
of the spin coherence, limiting the storage time to a few milliseconds
\cite{Buchler-efficiency-storageTime-review,Novikova-review}. Here
we present and experimentally demonstrate a storage scheme that is
insensitive to spin-exchange collisions, thus enabling long storage
times at high atomic densities. This unique property is achieved by
mapping the light field onto spin orientation within a decoherence-free
subspace of spin states. We report on a record storage time of 1 second
in cesium vapor, a 100-fold improvement over existing storage schemes.
Furthermore, our scheme lays the foundations for hour-long quantum
memories using rare-gas nuclear spins.}
\end{abstract}
\maketitle
The archetypal mechanism of light storage is based on electromagnetically
induced transparency (EIT), involving the \emph{signal} field to be
stored and a strong \emph{control} field \cite{Lukin-PRL-2001,Weitz-Spinor-polaritons,Novikova-integrated-gain,Halfmann-EIT-solid-state}.
These fields resonantly couple one atomic excited state to two spin
states within the ground level. Under EIT, there exists a long-lived
dark state, which is the superposition of these spin states that is
decoupled from the excited state due to destructive interference between
the excitation pathways. While the control is on, the signal pulse
entering the medium couples coherently to the dark state, forming
a slowly-propagating polariton. Storage is done by turning off the
control and stopping the polariton, thereby mapping the signal field
onto a stationary field of dark-state coherence. Turning on the control
retrieves the signal. 

\begin{figure}[tbh]
\begin{centering}
\includegraphics[bb=0bp 0bp 476bp 255bp,clip,width=7cm]{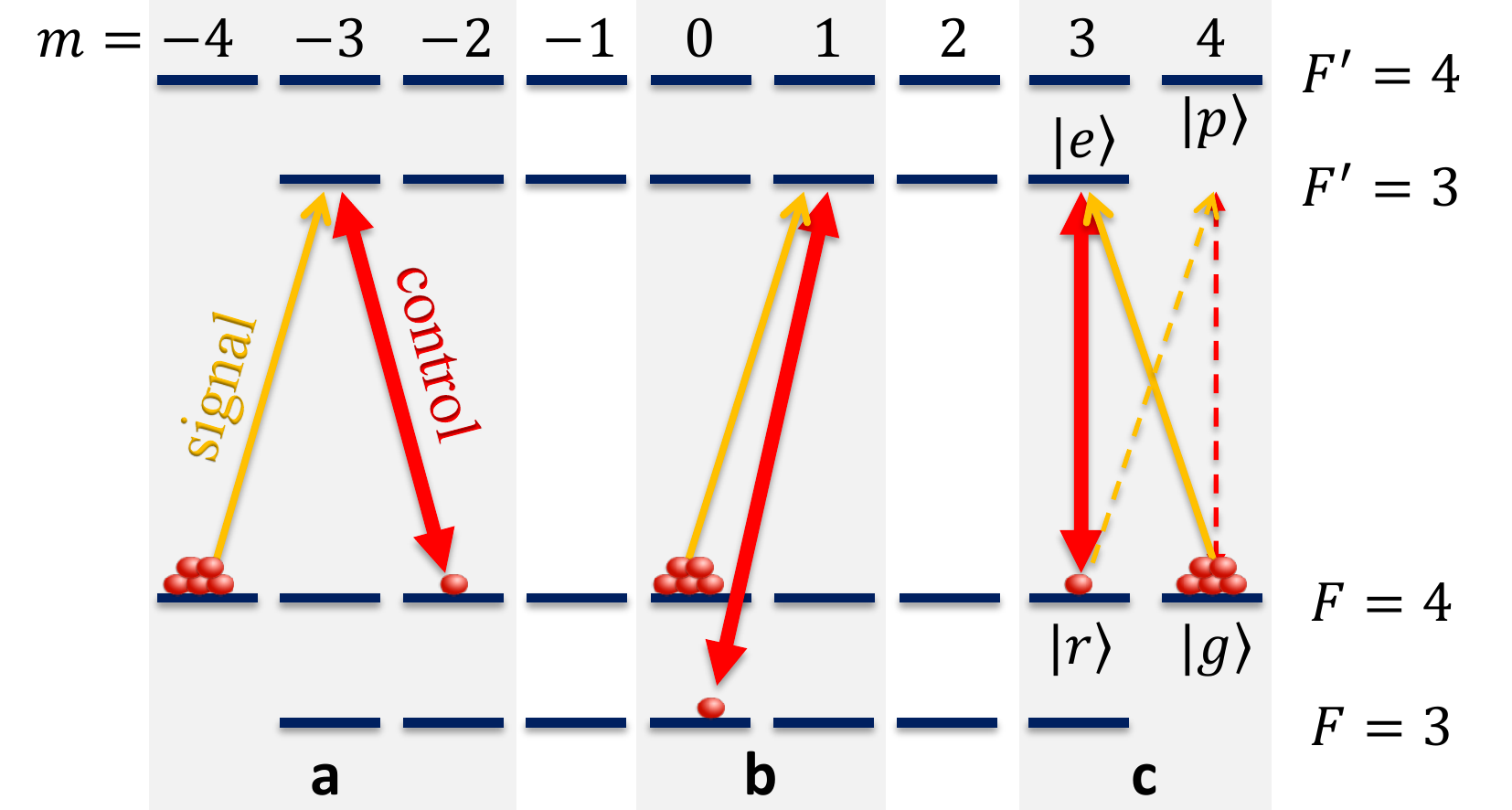}
\par\end{centering}
\caption{\textbf{Configurations of light storage on the cesium ground-level.}
\textbf{a}, Zeeman coherence $|\Delta m|=2$. \textbf{b}, Hyperfine
coherence $|\Delta m|=0$. \textbf{c}, Zeeman coherence $|\Delta m|=1$,
associated with spin orientation, used here. The dashed arrows represent
an additional weak process, discussed in the text. Note that the signal
is stored and retrieved as a linearly-polarized field, despite the
fact that only one of its circular-polarization components (solid
yellow line) enters the $\varLambda$-system $\left|g\right\rangle -\left|e\right\rangle -\left|r\right\rangle $,
see Methods. \label{fig: energy level}}
\end{figure}

In addition to the electron spin $S=1/2$, alkali-metal atoms have
a nuclear spin $I>0$ ($I=7/2$ for $^{133}$Cs) and thus possess
multiple spin states. These are characterized by the hyperfine spin
$F=I\pm S$ and its projection $m$ on the quantization axis $\hat{z}$.
Various combinations of spin states are accessible with different
signal-control configurations, as shown in Fig.~\ref{fig: energy level}.
Most light storage schemes utilize either the Zeeman coherence $\Delta m=2$
(Fig.~\ref{fig: energy level}a) or the hyperfine coherence $\Delta m=0$
(Fig.~\ref{fig: energy level}b) \cite{Novikova-review}. The relaxation
of these coherences at high atomic densities is dominated by pairwise
spin-exchange collisions \cite{Happer-1972}. During a collision,
the valence electrons of the colliding pair overlap for a few picoseconds,
accumulating a phase between the hybridized (singlet and triplet)
electronic spins. While the total spin is conserved, the randomness
of the collision parameters leads to relaxation of most ground-state
coherences, limiting the storage lifetime in these schemes \cite{Lukin-PRL-2001,Optimal-light-storage-Novikova-Gorshkov}.

It has been known, however, that the Zeeman coherence $\Delta m=1$,
associated with the spin orientation moment, is unaffected by spin-exchange
collisions at low magnetic fields \cite{Happre-SERF,Katz-nonlinear-SERF}.
This property is the underlying principle of spin-exchange relaxation-free
(SERF) magnetometers, currently the most sensitive magnetic sensors
\cite{Optical-magnetometery-Romalis,160-att-T-Romalis}. Here we realize
SERF light storage, closely related to the idea of a SERF atomic clock
\cite{Happer-clock}, by mapping the signal onto the $\Delta m=1$
coherence (Fig.~\ref{fig: energy level}c).

\begin{figure}[tbh]
\begin{centering}
\includegraphics[bb=0bp 0bp 567bp 522bp,clip,width=8cm]{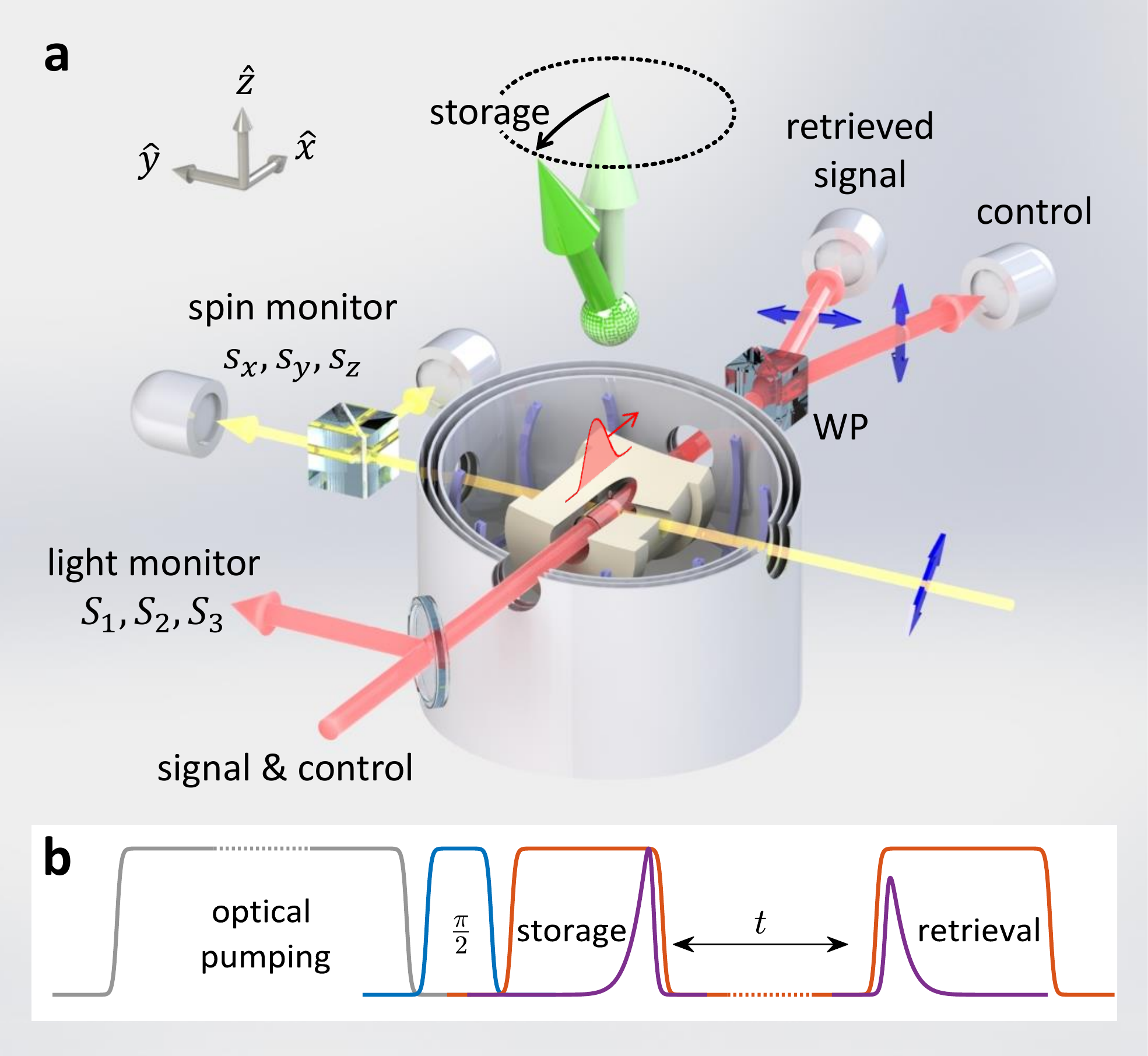}
\par\end{centering}
\caption{\textbf{Experimental setup and sequence.} A cylindrical vapor cell
with diameter 10 mm and length 30 mm is held inside a hot-air oven
(\textbf{a}, beige), enclosed by three Helmholtz coil pairs and magnetic
shield layers. First, optical pumping (\textbf{b}, gray) is done along
$\hat{x}$ using two circularly-polarized beams (not shown in \textbf{a})
for the two hyperfine sublevels. Subsequently, a $\pi/2$ pulse of
magnetic field along $\hat{y}$ (\textbf{b}, blue) prepares a spin
ensemble oriented along $\hat{z}$ (\textbf{a}, light green arrow\textbf{)}.
The control and incoming signal (\textbf{a}, red beam) are sampled
before the cell to monitor their polarization state. Storage and retrieval
of a signal pulse (\textbf{b}, purple) is done by turning off and
then on the control (\textbf{b}, red). The light state is stored onto
the orientation of the tilted spin polarization (\textbf{a}, dark
green arrow), which can be monitored with auxiliary far-detuned light
(\textbf{a}, yellow beam). The retrieved signal is separated from
the control using a high extinction-ratio Wollaston prism (WP). Blue
arrows in (\textbf{a}) indicate optical polarizations. \label{fig: exp_setup}}
\end{figure}

A paraffin-coated vapor cell is at the heart of the experimental system,
shown in Fig.~\ref{fig: exp_setup}a. We zero the magnetic field
to better than $\left|B\right|<1\,\mu G$ and control the cesium density
$n(T)$ via the cell temperature $T$. The experimental sequence is
shown in Fig.~\ref{fig: exp_setup}b. We initially orient the atoms
along the optical axis $\hat{x}$ using optical pumping (polarization
>95\%). We then rotate the polarized spins onto our quantization axis
$\hat{z}$ using a short pulse of magnetic field along $\hat{y}$,
thus preparing them in the state $\left|g\right\rangle \equiv\left|F=4,m=4\right\rangle $.
Subsequently, we turn on the control field and a small magnetic field
$B_{z}\leq15\,\mu$G. The control field is linearly polarized along
$\hat{z}$ and resonant with the $\left|r\right\rangle =\left|F=4,m=3\right\rangle \rightarrow\left|e\right\rangle =\left|F'=3,m'=3\right\rangle $
transition. 

\begin{figure}[tb]
\begin{centering}
\includegraphics[bb=0bp 0bp 372bp 278bp,clip,width=8cm]{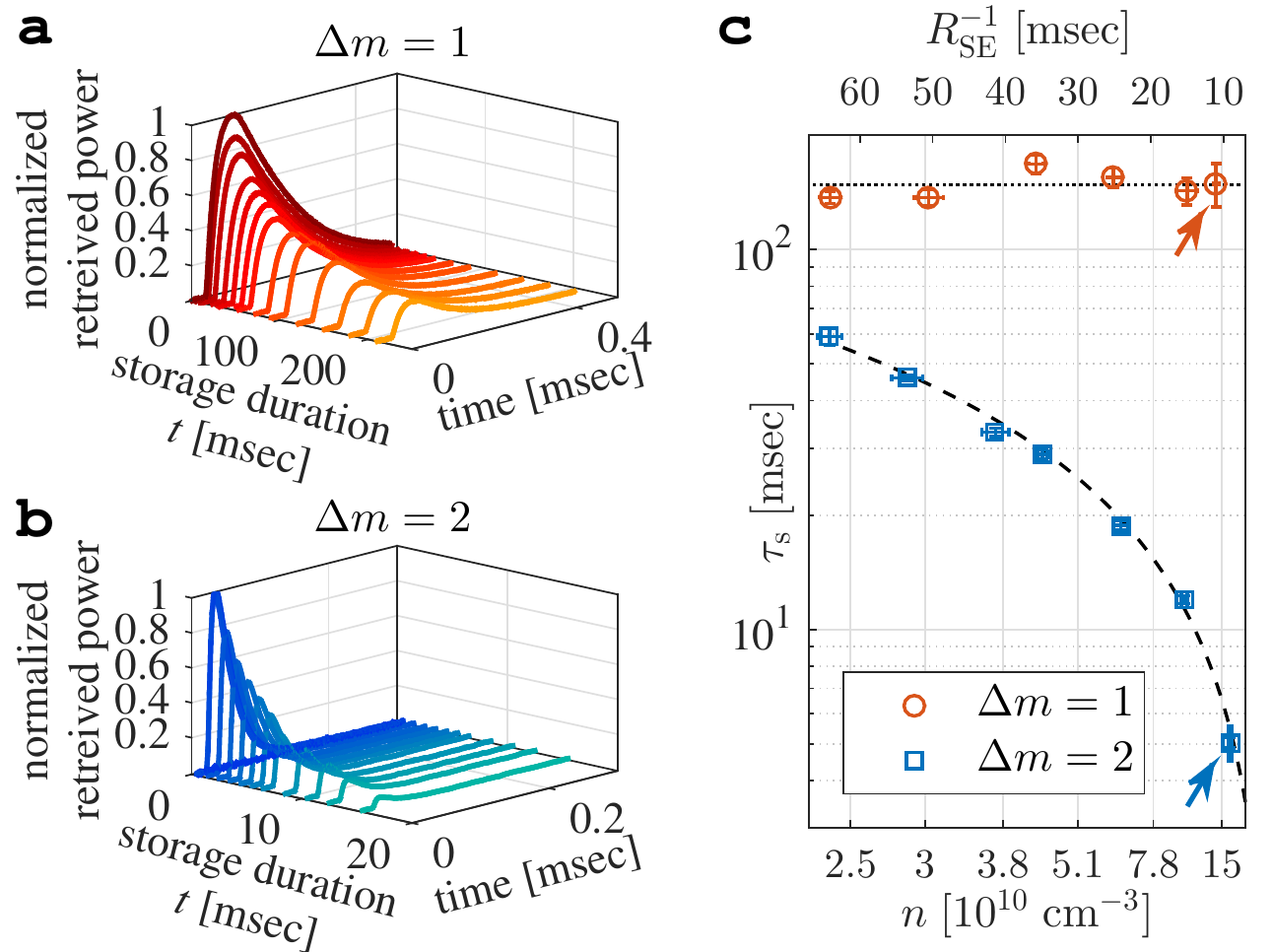}
\par\end{centering}
\caption{\textbf{Storage lifetime. a},\textbf{b}, Retrieved pulses for several
storage durations in our scheme $\Delta m=1$ (\textbf{a}), compared
to the standard scheme $\Delta m=2$ (\textbf{b}). The lifetime in
our scheme is longer than 100 msec, compared to only a few msec in
the standard scheme. \textbf{c}, Storage lifetime as a function of
atomic density $n$ and spin-exchange rate $R_{\mathrm{SE}}$ in our
scheme (red), which is unaffected by collisions and thus remains constant
(dotted line). In contrast, the lifetime of the standard scheme (blue)
is well described by a linear fit (dashed line). Data in (\textbf{a,b})
\textcolor{black}{correspond to points marked by arrows in} (\textbf{c}).
The two schemes exhibit comparable storage lifetimes only at low atomic
densities, where the optical depth and thus the storage efficiency
are compromised \cite{Lukin-RMP-2003}. \label{fig: retireved pulses}}
\end{figure}

With the control on, we send a weak signal pulse, linearly polarized
along $\hat{y}$. The signal couples to the $\left|g\right\rangle -\left|r\right\rangle $
coherence, orienting the spins while propagating. We store the signal
field onto spin orientation by turning off the control and, after
a duration $t$, retrieve it by turning the control back on. As a
reference, we perform light storage in a standard $\Delta m=2$ scheme
\cite{Lukin-PRL-2001}. At a density of $n=1.4\times10^{11}\;\mathrm{cm}^{-3}$
($T\approx40^{\circ}$ C), the two schemes exhibit comparable (internal)
storage efficiency, on order 10\%, with no particular optimization
of the temporal shape of the control and signal \cite{Optimal-light-storage-Novikova-Gorshkov}.
Figures~\ref{fig: retireved pulses}a,b show the retrieved pulses
for both schemes. We extract the storage lifetime $\tau_{\mathrm{s}}$
by fitting the retrieved power to the decay function $\exp(-t/\tau_{\mathrm{s}})$.
Light storage on spin orientation exhibits a remarkable lifetime $\tau_{\mathrm{s}}=149(20)$
msec in this experiment, much longer than the $5.0(3)$ msec obtained
with the standard scheme. 

To study the effect of spin-exchange collisions, we tune the collision
rate $R_{\mathrm{SE}}=\alpha n(T)$ by changing $T$ ($\alpha=6.5\cdot10^{-10}\,\mathrm{cm^{3}/}\mbox{sec}$
near room temperature) \cite{optically-pumped-atoms-book}. Figure~\ref{fig: retireved pulses}c
shows the measured storage lifetime versus $R_{\mathrm{SE}}$. The
relaxation of the $\Delta m=2$ coherence is dominated by spin exchange,
as indicated by the linear dependence of $\tau_{\mathrm{s}}$ on $R_{\mathrm{SE}}$
in the standard scheme. In contrast, our scheme is found to be insensitive
to $R_{\mathrm{SE}}$, affirming that the $\Delta m=1$ coherence
is conserved under spin exchange. We conclude that storage on spin
orientation maintains long memory lifetimes at elevated optical depths.
The observed lifetime $\tau_{\mathrm{s}}=150$ msec is limited by
the spin-destruction time at low magnetic fields, measured $T_{1}=300\pm100$
msec in our system. 

\begin{figure}[t]
\begin{centering}
\includegraphics[width=8cm]{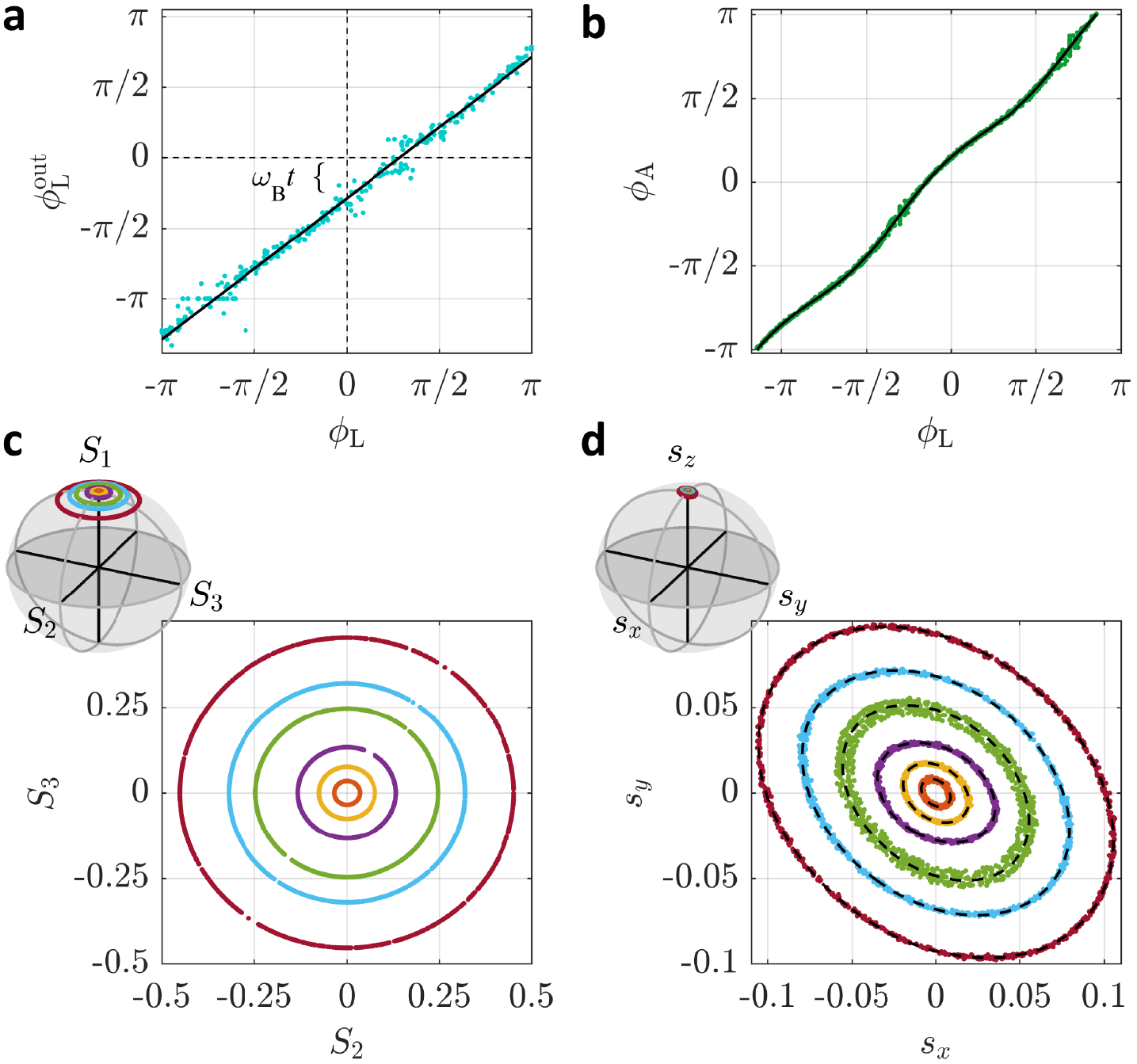}
\par\end{centering}
\caption{\textbf{Mapping from light polarization to spin orientation and back.
a}, The measured phase of the retrieved light after storage for $t=100$
msec follows the input phase up to an offset (line is a linear fit
with unity slope). \textbf{b}, The azimuthal angle of the collective
atomic spin versus the incoming optical phase (line is a fit to elliptical
mapping). \textbf{c}, Light polarization visualized on the Poincar\'e
sphere and projected onto the transverse plane. \textbf{d}, Collective
atomic spin visualized on the Bloch sphere and projected onto the
transverse plane. Data in \textbf{b-d} taken 1 msec after storage
(without retrieval). \label{fig: Tomography} }
\end{figure}

We confirm the coherent nature of our storage scheme by measuring
for $t=100\;\mbox{msec}$ the phases of the input signal $\phi_{\mathrm{L}}$
and output signal $\phi_{\mathrm{L}}^{\mathrm{out}}$, as shown in
Fig.~\ref{fig: Tomography}a. Larmor precession during storage leads
to a constant offset between $\phi_{\mathrm{L}}^{\mathrm{out}}$ and
$\phi_{\mathrm{L}}$. The Larmor frequency in this experiment was
measured\textcolor{black}{{} independently to be $\omega_{B}=1.34(6)\cdot2\pi$
Hz, predicting a rotation of $\omega_{\mathrm{B}}t=0.84(4)$, in agreement
with the observed offset $\phi_{\mathrm{L}}-\phi_{\mathrm{L}}^{\mathrm{out}}=0.9(2)$. }

To explain the immunity to spin-exchange collisions, we first explore
the light-atom mapping. Taking the control field as a phase reference,
the signal properties, or the light 'state' to be stored, are encompassed
in the polarization of the incoming (signal+control) field. This polarization
is visualized on the Poincar\'e sphere using the Stokes parameters
$S_{1},S_{2},S_{3}$ in Fig.~\ref{fig: Tomography}c. To characterize
their mapping onto the atomic spins, we monitor the spins during storage
using polarization rotation of a far-detuned beam. The spin state
of the ensemble is described by the collective electronic spin $\vec{s}=(s_{x},s_{y},s_{z})$,
defined by $\vec{s}=\frac{1}{N}\sum_{i}\langle\vec{s}^{i}\rangle$,
where $\vec{s}^{i}$ is the spin operator of the $i^{\mathrm{th}}$
atom and $N$ the number of atoms \cite{Polzik-2002}. These are visualized
on the Bloch sphere in Fig.~\ref{fig: Tomography}d. 

From the Stokes operators, we extract for the light the complex ratio
$i\eta_{\mathrm{L}}e^{i\phi_{\mathrm{L}}}=E_{\mathrm{s}}/E_{\mathrm{c}}$
between the signal amplitude $E_{\mathrm{s}}$ and the control amplitude
$E_{\mathrm{c}}$. The phase $\phi_{\mathrm{L}}$ is constant in space
and time during an experimental sequence. As the signal is much weaker
than the control $\eta_{\mathrm{L}}\ll1$, the Stokes vector is located
near the north pole of the Poincar\'e sphere (Fig.~\ref{fig: Tomography}c).
Initially, with only the control on ($\eta_{\mathrm{L}}=0$), the
atomic spins are oriented along the $\hat{z}$ direction, corresponding
to the north pole of the Bloch sphere. This initialized spin orientation
underlies the difference between our system and those previously demonstrated
with the same fields configuration \cite{Weitz-Spinor-polaritons,Novikova-review,Goldfarb-CPO}.
The incoming signal tilts the collective spin off the pole, producing
transverse spin components (Fig.~\ref{fig: Tomography}d). 

The light state has a polar angle $\eta_{\mathrm{L}}$ and an azimuth
$\phi_{\mathrm{L}}$; the corresponding atomic state has\textcolor{black}{{}
a polar angle }$\eta_{\mathrm{A}}=\sqrt{s_{x}^{2}+s_{y}^{2}}/s_{z}$\textcolor{black}{{}
and az}imuth $\phi_{\mathrm{A}}=\arctan(s_{y}/s_{x})$. We find that
the storage procedure maps the light quadratures $S_{2},S_{3}$ onto
the spin components $s_{y},s_{x}$ by transforming circles into nearly-circular
ellipses with $\phi_{\mathrm{A}}\approx\phi_{\mathrm{L}}$, see Fig.~\ref{fig: Tomography}b.
We conclude that the light state is mapped onto the atomic spin orientation.

The immunity to spin-exchange collisions can easily be understood
in the absence of hyperfine interaction (\emph{e.g.}~if $I=0$),
as the total spin, and hence the orientation of the electronic spin,
is conserved under these collisions. However when $I>0$, it is not
trivial to see how also the entanglement between the electronic and
nuclear spins is conserved. To understand the case $I>0$, we examine
a collision between two cesium atoms. After a weak classical signal
is stored, the state of the $i^{\mathrm{th}}$ atom is $\left|\psi_{i}\right\rangle =\left|g_{i}\right\rangle +\sqrt{2}\eta_{\mathrm{A}}e^{i\phi_{\mathrm{A}}}\left|r_{i}\right\rangle $,
with $\eta_{\mathrm{A}}\ll1$. The collision is brief relatively to
the hyperfine frequency and thus affects only the electronic spins.
We therefore decompose the stored state into the electronic $\left|\uparrow_{i}\right\rangle \equiv\left|s_{z}^{i}=\frac{1}{2}\right\rangle $,
$\left|\downarrow_{i}\right\rangle \equiv\left|s_{z}^{i}=-\frac{1}{2}\right\rangle $
and nuclear $\left|\Uparrow_{i}\right\rangle \equiv\left|I_{z}^{i}=I\right\rangle $,
$\left|\Downarrow_{i}\right\rangle \equiv\left|I_{z}^{i}=I-1\right\rangle $
spin components, writing $\left|\psi_{i}\right\rangle =\left|\uparrow_{i}\Uparrow_{i}\right\rangle +\sqrt{2}\eta_{\mathrm{A}}e^{i\phi_{\mathrm{A}}}\left(q\left|\uparrow_{i}\Downarrow_{i}\right\rangle +p\left|\downarrow_{i}\Uparrow_{i}\right\rangle \right)$.
For cesium, $I=\frac{7}{2}$ and $p^{2}=1-q^{2}=\frac{1}{8}$ (note
that the following arguments are general and independent of these
values). The exchange interaction during a collision between a pair
of atoms introduces a random phase $\chi$ between their hybrid electronic
states \textendash{} the singlet and triplet \cite{Happer-1972}.
The pair, initially in the product state $\left|\psi_{ij}\right\rangle =\left|\psi_{i}\right\rangle \left|\psi_{j}\right\rangle $,
leaves the collision in the state $(P_{T}^{i,j}+e^{i\chi}P_{S}^{i,j})\left|\psi_{ij}\right\rangle $,
where $P_{S}^{i,j}=(\left|\uparrow_{i}\downarrow_{j}\right\rangle -\left|\downarrow_{i}\uparrow_{j}\right\rangle )(\left\langle \uparrow_{i}\downarrow_{j}\right|-\left\langle \downarrow_{i}\uparrow_{j}\right|)/2$
and $P_{T}^{i,j}=1-P_{S}^{i,j}$ are the singlet and triplet projection
operators. Yet for weak signals, at the limit $\eta_{\mathrm{A}}\rightarrow0$,
the colliding pair is a nearly-perfect spin triplet, possessing a
negligible singlet component $\left\langle \psi_{ij}\right|P_{S}^{i,j}\left|\psi_{ij}\right\rangle =4(pq)^{2}\eta_{\mathrm{A}}^{4}\rightarrow0$
\cite{Happer-clock}. Therefore, the random phase $\chi$ is inconsequential,
and the pair state is immune to spin-exchange relaxation.

It is also instructive to examine the quantum limit, where the signal
has at most a single photon in the state $\alpha\left|0\right\rangle +\beta\left|1\right\rangle $,
where $\alpha$ and $\beta$ are the SU(2) parameters of a qubit with
either zero $\left|0\right\rangle $ or one $\left|1\right\rangle $
photons. At storage, the initial collective atomic state $\left|G\right\rangle =\prod_{i}\left|g_{i}\right\rangle $
is transformed into $\left|R\right\rangle =(\alpha+\beta F_{-})\left|G\right\rangle $,
where $F_{-}=\frac{1}{N}\sum_{i}(s_{-}^{i}+i_{-}^{i})$ is the collective
spin operator accounting for the $\Delta m=1$ transition. One can
verify that $\left|R\right\rangle $ is an exact triplet for any atom
pair ($i,j$), since $\left|G\right\rangle $ is a triplet and $[P_{T}^{i,j},F_{-}]=0$.
Therefore, the stored qubit $\left|R\right\rangle $ is fully conserved
under spin exchange. 

Finally, we note that a mapping $\eta_{\mathrm{L}}e^{i\phi_{\mathrm{L}}}\leftrightarrow\eta_{\mathrm{A}}e^{i\phi_{\mathrm{A}}}$
with nonzero ellipticity is non-ideal for quantum memories, as it
links the retrieval amplitude to the (azimuthal) phase and thus distorts
the quantum state. In our experiment, the ellipticity originates from
polarization self-rotation\textcolor{black}{{} \cite{Budker-Yavchuk-Self-Rotation-2001}}
due to the off-resonance Raman process $\left|g\right\rangle -\left|p\right\rangle -\left|r\right\rangle $
(dashed arrows in Fig.~\ref{fig: energy level}c) weakly perturbing
the ideal EIT process $\left|g\right\rangle -\left|e\right\rangle -\left|r\right\rangle $.
When the strength of these processes is comparable, the resulting
so-called Faraday interaction limits the storage to \textit{only}
one quadrature, compressing the ellipse into a line \cite{Polzik-2002}.
In our scheme, $\Delta>\Gamma$, where $\Delta$ and $\Gamma$ are
the detuning and Doppler-linewidth of the $\left|g\right\rangle -\left|p\right\rangle $
transition, so the dark state qualitatively obtains the form $\left|g\right\rangle +\sqrt{2}\eta_{\mathrm{L}}\left(e^{i\phi_{L}}-\epsilon e^{-i\phi_{L}}\right)\left|r\right\rangle $,
with $\epsilon=(1-i\Delta/\Gamma)^{-1}$; in cesium, $\Delta\approx10\Gamma$,
yielding ellipticity of order 0.1. An ideal mapping $\epsilon\rightarrow0$
with $\Delta\gg\Gamma$ is possible, \emph{e.g.}, by storing on the
lower hyperfine ground-level ($\left|r\right\rangle \Rightarrow\left|F=4,m=3\right\rangle $),
which still maintains the spin-exchange resistance. We derive in the
SI the exact analytical form of the mapping $\eta_{\mathrm{L}}e^{i\phi_{\mathrm{L}}}\leftrightarrow\eta_{\mathrm{A}}e^{i\phi_{\mathrm{A}}}$
and further develop a procedure employing a magnetic field $B_{z}$
that corrects for and eliminates the ellipticity. It follows that
the ellipticity is non-fundamental and amendable.

\begin{figure}[tb]
\centering{}\includegraphics[bb=0bp 0bp 410bp 197bp,clip,width=7cm]{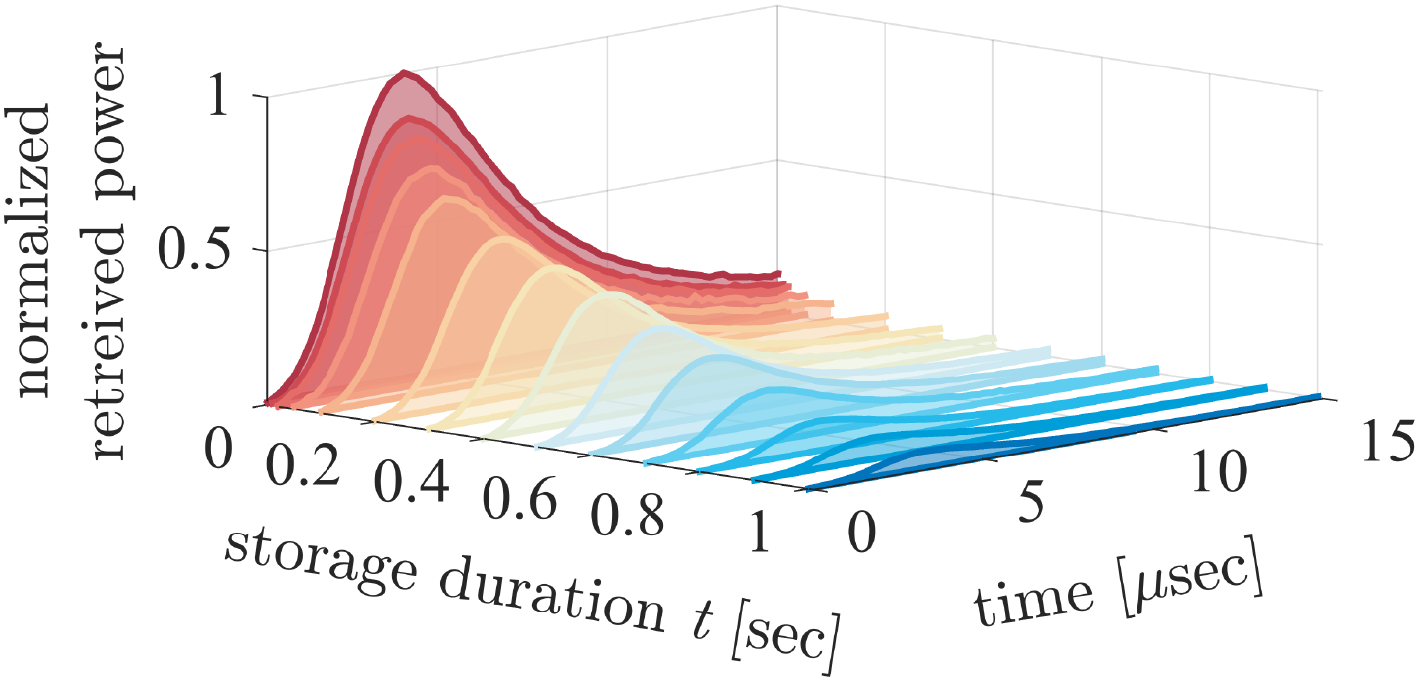}\caption{Light storage for up to $t=1\;\mbox{sec}$ of short ($\tau_{\mathrm{p}}=5.5\,\mu$sec)
signal pulses.\label{fig:onesecond} }
\end{figure}

Upon completion of the measurements reported above, we kept the vapor
cell warm at $T=45^{\circ}$C for a week, keeping the stem cold at
$T=25^{\circ}$C, and performed storage experiments for up to $t=1\;\mbox{sec}.$
As shown in Fig~\ref{fig:onesecond}, we observed a $1/e$ storage
time of $\tau_{\mathrm{s}}=430$(50) msec, indicating that the temperature
cycle lowered the spin destruction, presumably due to ``curing''
of the coating \cite{alkali_curing}. In conjunction with the signal
pulse duration of $\tau_{\mathrm{p}}=5.5\,\mu$sec, we thus obtained
an extremely large fractional delay of $\tau_{\mathrm{s}}/\tau_{\mathrm{p}}\approx80,000$.

In conclusion, our light-storage scheme demonstrates record lifetimes
of hundreds of msec at room temperature. In addition, it paves the
way towards the coupling of photons to ultra-stable nuclear spins.
Rare isotopes of noble gases, such as helium-3, carry nuclear spins
that are optically inaccessible and exhibit coherence times on the
scale of hours \cite{SEOP-Happer-RMP}. It has been demonstrated that
spin-exchange collisions can \emph{coherently} couple the orientation
moment of two alkali species \cite{Haroce-1970-coherent-coupling,katz-coherent-coupling}
as well as the orientation moment of alkali and rare-gas atoms \cite{Kornack-2002},
while all other spin moments are relaxed. Mapping light onto spin
\textit{orientation} thus not only protects the stored information
from self spin-exchange relaxation, but also enables its transfer
from one spin ensemble to another. Therefore, combined with coherent
spin-exchange interaction between alkali and rare gases, our scheme
is potentially a key element in employing and manipulating nuclear
spins for quantum information applications.

\section*{Acknowledgments}

We thank O.~Peleg, C.~Avinadav, and R.~Shaham for helpful discussions
and M.~Sturm and P.~Fierlinger for providing us the vapor cell.
We acknowledge financial support by the Israel Science Foundation
and ICORE, the European Research Council starting investigator grant
Q-PHOTONICS 678674,  the Minerva Foundation, the Sir Charles Clore
research prize, and the Laboratory in Memory of Leon and Blacky Broder.

\section*{Methods}

\emph{Experimental calibration.\textemdash{}} We perform calibration
experiments prior to storage, adjusting the setup parameters to produce
no output signal in the absence of an incoming signal. In particular,
we fine tune the spin rotation from $\hat{x}$ to $\hat{z}$ after
optical pumping and zero the transverse magnetic fields ($B_{x}$
and $B_{y}$). This calibration is important, as light storage based
on $|\Delta m|=1$ coherence is particularly sensitive to experimental
imperfections affecting the collective spin orientation: Nonzero \textit{\emph{transverse
magnetic fields}} ($B_{x},B_{y}\ne0$) tilt the collective spin during
storage and produce a small transverse spin component, which is subsequently
mapped to an output signal even without an input signal. Additionally,
misalignment\textit{\emph{ between the }}direction of the (linear)
polarization of the control field and the initial polarization direction
of the spin ensemble manifests as a nonzero transverse spin (when
identifying the control polarization as the quantization axis), again
producing an output signal for no input signal. We thus properly align
the initial spin polarization direction and validate that $B_{x}$
and $B_{y}$ are truly zeroed by verifying the absence of output signal
for all ``storage'' durations $t$, confirming that there is absolutely
no tilt of the spin during the experiment.

\emph{Phase uniformity.\textemdash{} }In the storage experiments,
we send weak signal pulses of durations $\tau_{\mathrm{p}}=0.03-0.15$
ms, linearly polarized along $\hat{y}$, and having the same spatial
mode and frequency as the control field. The corresponding range of
pulse bandwidths $2\tau_{\mathrm{p}}^{-1}\approx2-10\cdot2\pi$ kHz
is comparable to the width of the EIT transmission window and much
larger than the Larmor precession rate $\omega_{\text{B}}$, such
that the relative phase between the signal and control fields is constant
during a storage experiment. The signal and control beams cover the
entire cell volume, with their wave-vector difference much smaller
than the inverse cell width. This was chosen, because the storage
lifetime is sensitive to the \textit{\emph{spatial-mode overlap}}
of the signal and control fields. Specifically, an angular deviation
between them yields a spatial phase grating, which is imprinted on
the collective spin wave. Dephasing of this spin wave due to thermal
atomic motion limits the storage lifetime \cite{Firstenberg-residual-doppler},
as it does regardless of the exact spin coherence used. However for
the $|\Delta m|=1$ scheme, the spin-wave grating manifests as a spatially
varying orientation, impairing the resistance to spin-exchange collisions.
Colliding spins with different orientations are no longer perfect
triplets (the singlet component $\left\langle \psi_{ij}\right|P_{S}^{i,j}\left|\psi_{ij}\right\rangle $
grows quadratically with the angle between the colliding spins), which
can be explained by their reduced indistinguishability due to the
spatially-varying mapping. Consequently, the $|\Delta m|=1$ scheme
would lose its SERF property were the signal and control beams misaligned. 

\emph{Completeness of the $|\Delta m|=1$ transition.\textemdash{}}
It is instructive to discuss an intricacy that arise when the quantization
axis ($\hat{z}$) is orthogonal to the light propagation axis ($\hat{x}$)
in the special case of an ensemble initially polarized (oriented)
along $\hat{z}$. With a control field linearly-polarized along $\hat{z}$,
the maximally-polarized atoms are confined to a single $|\Delta m|=1$
transition. From the viewpoint of the quantization axis $\hat{z}$,
this transition corresponds to the $\varLambda$-system $\left|g\right\rangle -\left|e\right\rangle -\left|r\right\rangle $,
as depicted in Fig.~\ref{fig: energy level}. Importantly, the signal
mode is completely stored and retrieved via this transition, despite
the fact that its linear polarization ($\hat{y}$) decomposes to two
circular polarizations, of which only one is included in the $\left|g\right\rangle -\left|e\right\rangle -\left|r\right\rangle $
system. Under the reduced Maxwell equations, including the transverse
susceptibility tensor of the medium \cite{optically-pumped-atoms-book},
the normal modes comprises only $E_{y}$ and $E_{z}$ components (recall
that the non-evanescent field polarization remains within the transverse
plane during praxial propagation). Consequently, the signal mode $E_{y}$
is stored and retrieved 'as a whole'. The optical depth for the signal
is however reduced by the Clebsch-Gordan coefficient of the $\left|g\right\rangle -\left|e\right\rangle $
transition. 

\emph{Measuring light and spin polarizations.\textemdash{}} To measure
the Stokes parameters of the light and extract $\eta_{\mathrm{L}}$
and $\phi_{\mathrm{L}}$, we sample the optical fields before the
cell. To measure the collective atomic spin, we use a weak monitor
beam propagating along $\hat{y}$, linearly polarized ($\hat{z}$)
and red-detuned $22\;\mbox{GHz}$ from the $\left|g\right\rangle \rightarrow\left|e\right\rangle $
transition. Far from resonance, the polarized atoms render the medium
optically chiral, rotating the polarization of the monitor beam in
the $xz$ plane by an angle $\theta=\beta s_{y}$ via the linear Faraday
interaction, where $\beta$ is a constant \cite{optically-pumped-atoms-book}.
We measure $\theta$ after the cell using a balanced detector \cite{katz-coherent-coupling}.
The collective spin during storage is measured by increasing the magnetic
field to $B_{z}=4$ mG, making the spin precess around the $\hat{z}$
axis at a frequency $\omega_{\mathrm{B}}=1.4\cdot2\pi$ kHz and thus
modulating $\theta$ in time according to $\theta=C\cos\left(\omega_{\mathrm{B}}t+\phi_{\mathrm{A}}\right)$.
We identify the transverse spin components at storage as $s_{x}=C\cos\left(\phi_{\mathrm{A}}\right)/\beta$
and $s_{y}=C\sin\left(\phi_{\mathrm{A}}\right)/\beta$. We scan the
input phase $\phi_{\mathrm{L}}$ at various signal powers and measure
$s_{x}$, $s_{y}$ in each realization. We perform an additional set
of experiments by applying $B_{x}$ instead of $B_{z}$, thus modulating
the spin in the $yz$ plane and measuring $\overline{s_{z}}$ averaged
per signal power. With $s_{x}$, $s_{y}$, and $\overline{s_{z}}$,
we extract $\eta_{\mathrm{A}}=\sqrt{s_{x}^{2}+s_{y}^{2}}/\overline{s_{z}}$
and $\phi_{\mathrm{A}}=\arctan(s_{y}/s_{x})$. For each measurement,
we use the normalized vector $\vec{s}/\left\Vert \vec{s}\right\Vert $
to lay the spin on the Bloch sphere and eliminate $\beta$, which
is independent of the signal parameters\textcolor{magenta}{{} }\cite{optically-pumped-atoms-book}.

\onecolumngrid \appendix 
\setcounter{equation}{0}
\setcounter{figure}{0}
\setcounter{table}{0}
\setcounter{page}{1}
\makeatletter
\renewcommand{\theequation}{S\arabic{equation}}
\renewcommand{\thefigure}{S\arabic{figure}}
\renewcommand{\bibnumfmt}[1]{[S#1]}
\renewcommand{\citenumfont}[1]{S#1} 

\part*{Supplementary Information for ``Light storage for one second at
room temperature''}

In this Supplementary Information, we analytically derive the mapping
from light to atoms $\eta_{\mathrm{L}}e^{i\phi_{\mathrm{L}}}\rightarrow\eta_{\mathrm{A}}e^{i\phi_{\mathrm{A}}}$
(Sec.~I) and from atoms to light $\eta_{\mathrm{A}}e^{i\phi_{\mathrm{A}}}\rightarrow\eta_{\mathrm{L}}^{\mathrm{out}}e^{i\phi_{\mathrm{L}}^{\mathrm{out}}}$(Sec.~II).
Subsequently, we examine the overall mapping between the input and
output signals and present a method to eliminate its ellipticity (Sec.~III). 

\subsection*{I. Storage: mapping light to atoms}

In light storage based on EIT, the atomic state during storage corresponds
to the EIT dark state. Here we derive the dark state by analyzing
the non-Hermitian Hamiltonian of the system \cite{footnote1}. For
our $\Delta m=1$ scheme, the control field $E_{\mathrm{c}}$ and
signal field $E_{\mathrm{s}}$ comprise the total electric field as
\begin{equation}
\vec{E}=E_{\mathrm{c}}\hat{z}+E_{\mathrm{s}}\hat{y}=E_{c}\left(\hat{z}+i\eta_{\mathrm{L}}e^{i\phi_{\mathrm{L}}}\hat{y}\right).
\end{equation}
In principle, the ground level of the Cs atoms have many spin states
that are coupled by the signal and control fields. However, we use
optical pumping to initialize the atoms in the maximally polarized
state $\left|g\right\rangle $ (see Fig.~1 in the main text), while
the weak signal field with $\eta_{\mathrm{L}}\ll1$ varies this state
only perturbatively. We therefore consider here, in addition to $\left|g\right\rangle $,
only the states $\left|r\right\rangle $,$\left|e\right\rangle $,$\left|p\right\rangle $,
which couple to $\left|g\right\rangle $ to first order in $\eta_{\mathrm{L}}$.
The non-Hermitian Hamiltonian is given by $H=H_{0}+V$, where

\begin{align}
H_{0} & =-i\Gamma\left|e\right\rangle \left\langle e\right|+\left(\Delta-i\Gamma\right)\left|p\right\rangle \left\langle p\right|\\
V & =\Omega_{s}\left|g\right\rangle \left\langle e\right|+\Omega_{c}\left|r\right\rangle \left\langle e\right|+a_{{\scriptscriptstyle \textsc{\textnormal{Cs}}}}\Omega_{c}\left|g\right\rangle \left\langle p\right|+b_{{\scriptscriptstyle \textsc{\textnormal{Cs}}}}\Omega_{s}\left|r\right\rangle \left\langle p\right|+\mbox{h.c.}\nonumber 
\end{align}
Here $\Delta=1100\cdot2\pi\;\mbox{MHz}$ is the excited-level hyperfine
splitting, and we take $\Gamma=124\cdot2\pi\;\mbox{MHz}$ for the
half linewidth of the Doppler-broadened optical transition. The Rabi
frequencies are given by $\Omega_{c}=d_{{\scriptscriptstyle \textsc{\textnormal{Cs}}}}\left|E_{\mathrm{c}}\right|/(\sqrt{2}\hbar)$
and $\Omega_{s}=\sqrt{2}\Omega_{c}\eta_{\mathrm{L}}e^{i\phi_{L}}$,
where $d_{{\scriptscriptstyle \textsc{\textnormal{Cs}}}}=2.6\cdot\frac{\sqrt{7}}{4}ea_{0}$
is the\textcolor{black}{{} dipole moment transition element for the
Cs D1 transition}, with the electron charge $e$ and Bohr radius $a_{0}$
\cite{Steck-Cs-SI}. The ratios of the Clebsch-Gordan coefficients
between the two $\Lambda$ systems is $a_{{\scriptscriptstyle \textsc{\textnormal{Cs}}}}=4/\sqrt{7}$
and $b_{{\scriptscriptstyle \textsc{\textnormal{Cs}}}}=1/\sqrt{7}$.

\begin{figure}[h]
\begin{centering}
\includegraphics[width=14cm]{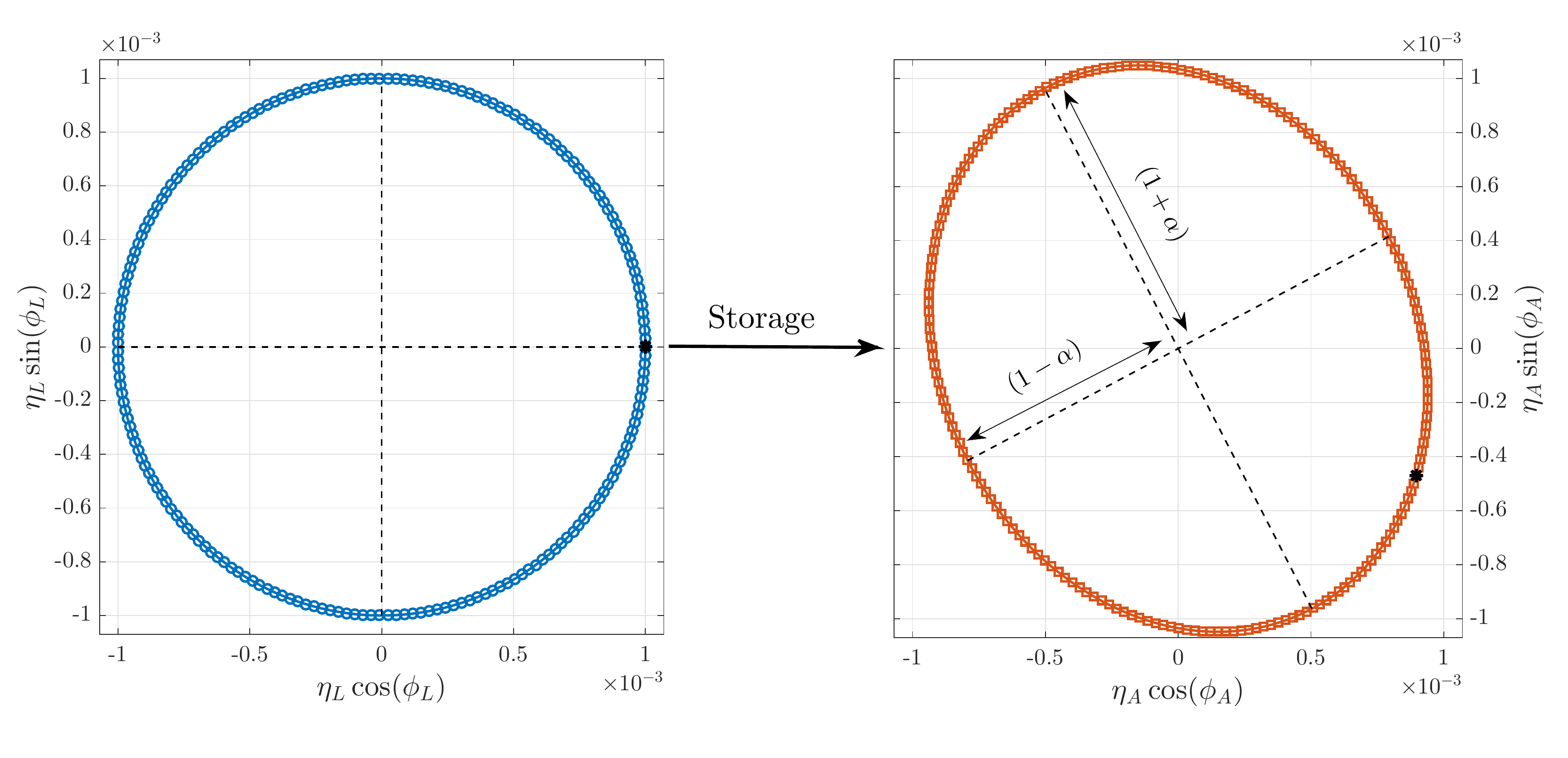}
\par\end{centering}
\caption{Mapping of the light state to atomic spin state ($\eta_{\mathrm{L}}=10^{-3}$,
$\alpha=0.1$). We mark the points $\phi_{\mathrm{L}}=0$ and $\phi_{\mathrm{A}}$$\left(\phi_{\mathrm{L}}=0\right)$
(black asterisks) to illustrate the transformation of the azimuthal
phase, which acquires a mean additional phase shift of $-3\alpha$.
\label{fig: storage mapping analytical}}
\end{figure}

Because of the off-resonant coupling to the state $\left|p\right\rangle $,
the system has no dark state, as manifested by the fact that $H$
has no zero eigenvalues. However, one can identify a ``quasi-dark''
state \textemdash{} the eigenstate of $H$ with the lowest imaginary
eigenvalue. For our Hamiltonian, the lowest imaginary eigenvalue $\lambda_{\mathrm{min}}\propto-i\Gamma\Omega_{c}^{2}/\Delta^{2}$
accounts for the loss of population from the dark state to other ground-level
states via off-resonant pumping. Diagonalizing $H$, we find the corresponding
eigenstate to first order in $\eta_{\mathrm{L}}$ and in $\Gamma/\Delta$, 

\begin{equation}
\left|\psi_{\mathrm{A}}\right\rangle =\left|g\right\rangle +\sqrt{2}\eta_{\mathrm{L}}e^{-3i\alpha}[e^{i\left(\phi_{\mathrm{L}}-\alpha\right)}\underbrace{-i\alpha e^{-i\left(\phi_{\mathrm{L}}-\alpha\right)}}_{\mathrm{\begin{array}{c}
\mathrm{perturbation\,to\,the}\\
\mathrm{ideal\,dark\,state}
\end{array}}}]\left|r\right\rangle \label{eq:Atomic dark state}
\end{equation}
with $\alpha=f_{{\scriptscriptstyle \textsc{\textnormal{Cs}}}}\Gamma/\Delta$,
and $f_{{\scriptscriptstyle \textsc{\textnormal{Cs}}}}=0.58$ is a
constant derived from the Clebsch-Gordan coefficients. To relate this
state to the Bloch sphere representation, we write it as
\begin{equation}
\left|\psi_{\mathrm{A}}\right\rangle =\left|g\right\rangle +\sqrt{2}\eta_{\mathrm{A}}e^{i\phi_{\mathrm{A}}}\left|r\right\rangle \label{eq:atomic_state}
\end{equation}
and identify the Bloch spin quadratures 
\[
s_{x}=\frac{1}{2}\eta_{\mathrm{A}}\cos\left(\phi_{\mathrm{A}}\right);\;\;s_{y}=\frac{1}{2}\eta_{\mathrm{A}}\sin\left(\phi_{\mathrm{A}}\right);\;\;s_{z}=\frac{1}{2}.
\]
The parameters $\eta_{\mathrm{A}}$ and $\phi_{\mathrm{A}}$ thus
serve as the angles on the Bloch sphere. With $\alpha\ne0$, the state
(\ref{eq:Atomic dark state}) manifests a non-ideal mapping from the
Poincar\'e sphere to the Bloch sphere. From Eqs.~(\ref{eq:atomic_state})
and (\ref{eq:Atomic dark state}), we find the $\eta_{\mathrm{L}}e^{i\phi_{\mathrm{L}}}\rightarrow\eta_{\mathrm{A}}e^{i\phi_{\mathrm{A}}}$
mapping

\begin{equation}
\eta_{\mathrm{A}}=\eta_{\mathrm{L}}\sqrt{1-2\alpha\sin\left(2\left(\phi_{\mathrm{L}}-\alpha\right)\right)}\label{eq: eta light 2 atoms mapping}
\end{equation}
and

\begin{equation}
\phi_{\mathrm{A}}=-\frac{\pi}{4}-3\alpha+\arctan\left[\frac{1-\alpha}{1+\alpha}\tan\left(\phi_{\mathrm{L}}+\frac{\pi}{4}-\alpha\right)\right].\label{eq: phi light 2 atoms mapping}
\end{equation}
These are the mathematical equations of an ellipse with the semi-major
and semi-minor axes $\eta_{\mathrm{L}}\left(1\pm\alpha\right)$ rotated
by an angle $\left(\frac{\pi}{4}-\alpha\right)$. Mathematically,
the phase $\phi_{\mathrm{L}}$ serves as the eccentric anomaly of
the ellipse in the Bloch sphere. The mapping is exemplified graphically
in Fig.~\ref{fig: storage mapping analytical} for $\eta_{\mathrm{L}}=10^{-3}$
and $\alpha=0.1$ with $0\leq\phi_{\mathrm{L}}\leq2\pi$. The ellipticity
of the mapping, determined by the parameter $\alpha,$ is manifested
by a dependence of the tilt angle of the spin on the azimuthal phase. 

As expected, in the ideal case $\alpha\rightarrow0$ (for $\Gamma\ll\Delta$),
the quasi-dark state in Eq.~(\ref{eq:Atomic dark state}) becomes
the ideal dark state, and Eqs.~(\ref{eq: eta light 2 atoms mapping})
and (\ref{eq: phi light 2 atoms mapping}) reduce to the simple linear
relations $\eta_{\mathrm{A}}=\eta_{\mathrm{L}}$ and $\phi_{\mathrm{A}}=\phi_{\mathrm{L}}$. 

\subsection*{II. Retrieval: mapping atoms to light}

\begin{figure}[th]
\begin{centering}
\includegraphics[width=14cm]{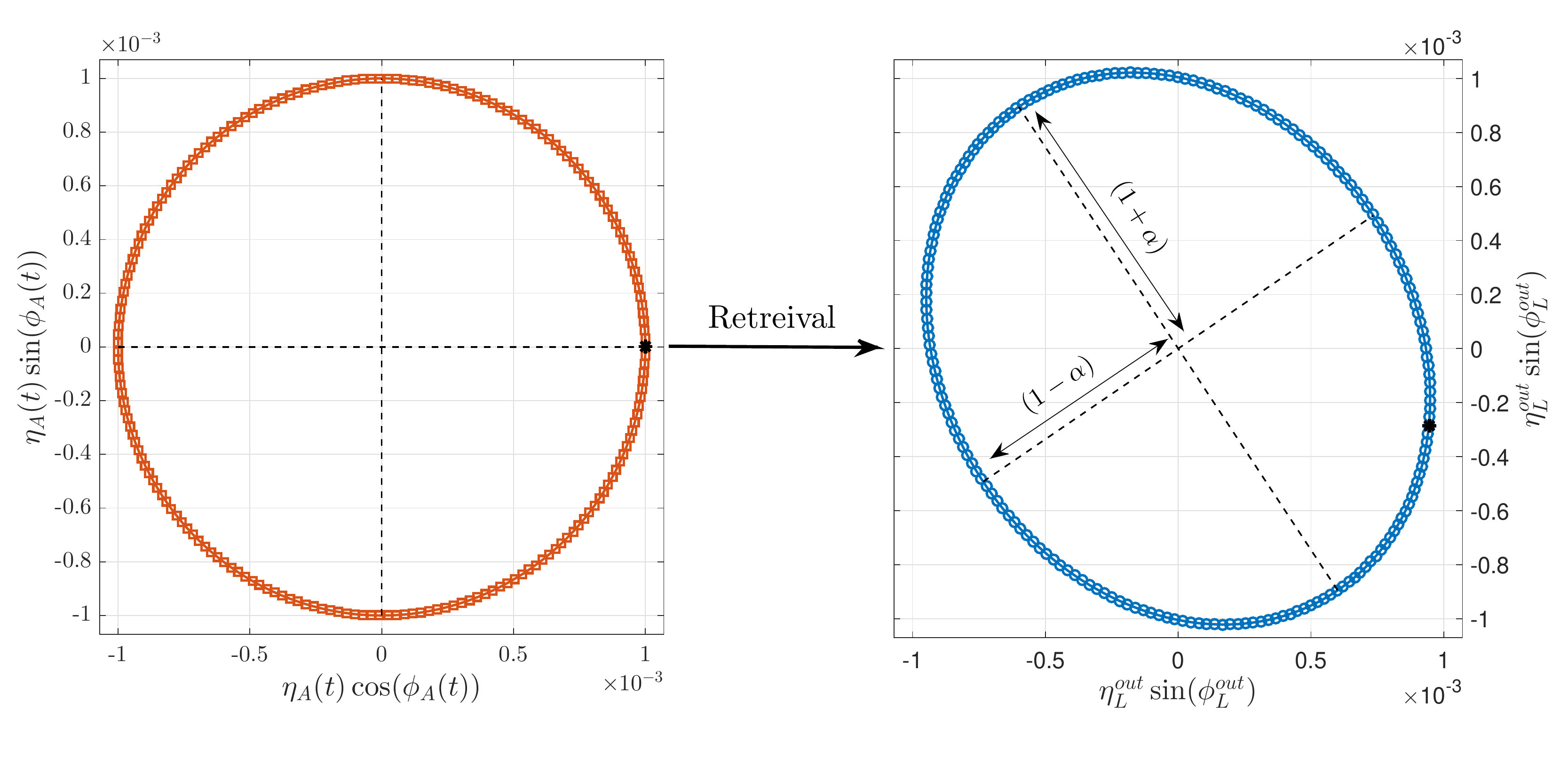}
\par\end{centering}
\caption{Mapping of the atomic spin state to the light state when retrieving
the signal ($\eta_{\mathrm{A}}=10^{-3}$, $\alpha=0.1$). We mark
the points $\phi_{\mathrm{A}}=0$ and $\phi_{L}\left(\phi_{\mathrm{A}}=0\right)$
(black asterisk) to illustrate the mapping of the azimuthal phase,
which acquires a mean additional phase shift of $-2\alpha$. \label{fig: retreival mapping analytical}}
\end{figure}

The retrieval process could be described as the reverse process of
storage \cite{Gorshkov-reversible,Nunn-reversible}. Nevertheless,
it is instructive to consider a complementary formalism, which we
give in this section. During the retrieval of the signal, similarly
to storage, the atomic excitation changes adiabatically the incoming
light field (control only) to a new light field (control+signal) that
minimizes the loss. A formalism for describing the propagation of
the electric field through the medium was introduced by Happer \textit{et
al.}~\cite{optically-pumped-atoms-book-SI}. They write the propagation
equation of the field amplitude $\vec{E}$ as 
\begin{equation}
\frac{d}{d\tilde{x}}\vec{E}=2\pi ikn\left\langle \overleftrightarrow{\chi}\right\rangle _{\bot}\vec{E},
\end{equation}
where $k$ is the wave-number of the laser, $n$ is the atomic density,
and $\tilde{x}=x-ct$ is the transformed spatial coordinate of the
pulse. This equation relates the electric field vector in the medium
$\vec{E}=(E_{y},E_{z})$ to the atomic state $\left|\psi_{\mathrm{A}}\right\rangle $
using the mean transverse susceptibility tensor of the atoms $\left\langle \overleftrightarrow{\chi}\right\rangle _{\bot}=\left\langle \psi_{\mathrm{A}}\left|\overleftrightarrow{\chi}\right|\psi_{\mathrm{A}}\right\rangle _{\bot}$,
where $\chi_{ij}=d_{i}d_{j}/(\Delta-i\Gamma)$ is the atomic susceptibility
operator, and $d_{i}$ are the atomic dipole operators. The subscript
``$\bot$'' denotes the reduced $2\times2$ operator, describing
the $yz$ polarization plane \cite{Happer-Tang-suceptability-1970}.
We identify the retrieved light field as the eigenvector of $\left\langle \overleftrightarrow{\chi}\right\rangle _{\bot}$
with minimal imaginary eigenvalue, that is, the light field with minimal
loss \cite{footnote2}. Using the atomic state (\ref{eq:atomic_state}),
the susceptibility tensor of the medium is given by

\begin{equation}
\left\langle \overleftrightarrow{\chi}\right\rangle _{\bot}=i\frac{d_{{\scriptscriptstyle \textsc{\textnormal{Cs}}}}^{2}}{\Gamma}\left(\begin{array}{cc}
1-2ib_{{\scriptscriptstyle \textsc{\textnormal{Cs}}}}^{2}\frac{\Gamma}{\Delta}\eta_{\mathrm{A}} & -i\eta_{\mathrm{A}}\left(e^{i\phi_{\mathrm{A}}}-ia_{{\scriptscriptstyle \textsc{\textnormal{Cs}}}}b_{{\scriptscriptstyle \textsc{\textnormal{Cs}}}}\frac{\Gamma}{\Delta}e^{-i\phi_{\mathrm{A}}}\right)\\
i\eta_{\mathrm{A}}\left(e^{-i\phi_{\mathrm{A}}}-ia_{{\scriptscriptstyle \textsc{\textnormal{Cs}}}}b_{{\scriptscriptstyle \textsc{\textnormal{Cs}}}}\frac{\Gamma}{\Delta}e^{i\phi_{\mathrm{A}}}\right) & \eta_{\mathrm{A}}-\frac{i}{2}a_{{\scriptscriptstyle \textsc{\textnormal{Cs}}}}^{2}\frac{\Gamma}{\Delta}
\end{array}\right).
\end{equation}
The least decaying eigenvector $\vec{E}^{\mathrm{out}}$ of this matrix
is given to first order in $\eta_{\mathrm{A}}$ and in $\Gamma/\Delta$
by

\begin{equation}
\vec{E}^{\mathrm{out}}=\left(\begin{array}{c}
i\eta_{\mathrm{A}}e^{-i2\alpha}\left(e^{i\phi_{\mathrm{A}}}-i\alpha e^{-i\phi_{\mathrm{A}}}\right)\\
1
\end{array}\right).\label{eq: light state}
\end{equation}
The resulting light field is elliptically polarized. We can relate
the polar and azimuthal angles on the  Poincar\'e sphere to those
of the Bloch sphere by 

\begin{equation}
\eta_{\mathrm{L}}^{\mathrm{out}}=\eta_{\mathrm{A}}\sqrt{1-2\alpha\sin\left(2\phi_{\mathrm{A}}\right)}\label{eq: eta atoms 2 light mapping}
\end{equation}
and

\begin{equation}
\phi_{\mathrm{L}}^{\mathrm{out}}=\frac{\pi}{4}-2\alpha+\arctan\left[\frac{1+\alpha}{1-\alpha}\tan\left(\phi_{\mathrm{A}}-\frac{\pi}{4}\right)\right].\label{eq: phi atoms 2 light mapping}
\end{equation}
Once again, these are the mathematical equations of an ellipse with
the semi-major and semi-minor axes $\eta_{\mathrm{L}}\left(1\mp\alpha\right)$,
now rotated by an angle $\frac{\pi}{4}$. The mapping is exemplified
graphically in Fig.~\ref{fig: retreival mapping analytical} for
$\eta_{\mathrm{A}}=10^{-3}$ and $\alpha=0.1$ with $0\leq\phi_{\mathrm{A}}\leq2\pi$.
The ellipticity of the mapping, determined by the parameter $\alpha,$
tilts the light polarization differently for different azimuthal phases.

Here again, in the ideal case $\alpha\rightarrow0$, the medium becomes
completely transparent to the state $\vec{E}^{\mathrm{out}}$, and
Eqs.~(\ref{eq: eta atoms 2 light mapping}) and (\ref{eq: phi light 2 atoms mapping})
reduce to the simple linear relations $\eta_{\mathrm{L}}^{\mathrm{out}}=\eta_{\mathrm{A}}$
and $\phi_{\mathrm{L}}^{\mathrm{out}}=\phi_{\mathrm{A}}$. 

\subsection*{III. Protocol for eliminating the ellipticity of the overall storage-to-retrieval
transformation}

The overall transformation of the storage, followed by a storage time
$t$, and then retrieval, is described by 

\begin{align}
\eta_{\mathrm{L}} & \rightarrow\eta_{\mathrm{A}}(0)\rightarrow\eta_{\mathrm{A}}(t)\rightarrow\eta_{\mathrm{L}}^{\mathrm{out}},\label{eq: total transfromation}\\
\phi_{\mathrm{L}} & \rightarrow\phi_{\mathrm{A}}(0)\rightarrow\phi_{\mathrm{A}}(t)\rightarrow\phi_{\mathrm{L}}^{\mathrm{out}}.\label{eq:total transformation phi}
\end{align}
The first and third steps are given by Eqs.~(\ref{eq: eta light 2 atoms mapping}),
(\ref{eq: phi light 2 atoms mapping}), (\ref{eq: eta atoms 2 light mapping}),
and (\ref{eq: phi atoms 2 light mapping}). The second step describes
the dynamics of the atomic spins in the dark for a duration $t$,
governed by a Larmor precession,

\begin{eqnarray}
\phi_{\mathrm{A}}(t) & = & \phi_{\mathrm{A}}(0)+\omega_{B}t,\label{eq: ground state phase}
\end{eqnarray}
and by decay due to the various relaxation mechanisms,

\[
\eta_{\mathrm{A}}(t)=\eta_{\mathrm{A}}(0)e^{-t/\tau_{\mathrm{s}}}.
\]

Ideally $\alpha=f_{{\scriptscriptstyle \textsc{\textnormal{Cs}}}}\Gamma/\Delta\ll1$
and $t\ll\tau_{\mathrm{s}}$, such that Eqs.~(\ref{eq: eta atoms 2 light mapping})
and (\ref{eq: phi atoms 2 light mapping}) are the inverse transformation
of Eqs.~(\ref{eq: eta light 2 atoms mapping}) and (\ref{eq: phi light 2 atoms mapping}),
and the ground state relaxation is negligible, yielding perfect storage
and retrieval of light. However, if $\alpha<1$ but non-negligible
(as in our experiment), then the retrieved signal is altered by the
coupling to the off-resonant level $\left|p\right\rangle $ and given
to first order in $\alpha$ by 

\begin{align}
\phi_{\mathrm{L}}^{\mathrm{out}} & \approx\phi_{\mathrm{L}}+\omega_{B}t-6\alpha-2\alpha\cos\left(\omega_{B}t-3\alpha\right)\cos\left(2\phi_{\mathrm{L}}+\omega_{B}t\right)\\
\eta_{\mathrm{L}}^{\mathrm{out}} & \approx\eta_{\mathrm{L}}e^{-t/\tau_{\mathrm{s}}}\left(1-2\alpha\sin\left(2\phi_{\mathrm{L}}+\omega_{B}t\right)\cos\left(\omega_{B}t-3\alpha\right)\right).
\end{align}
We see that, in general, the output light suffers the elliptical distortion
twice, resulting in a storage efficiency that depends on the phase.
This effect can be understood either as self rotation of the light
polarization \cite{novikova-gain-SI} or as arising from degenerate
four-wave mixing \cite{Xiao-phase-EIT-SI}. 

\begin{figure}[tbh]
\begin{centering}
\includegraphics[width=18cm]{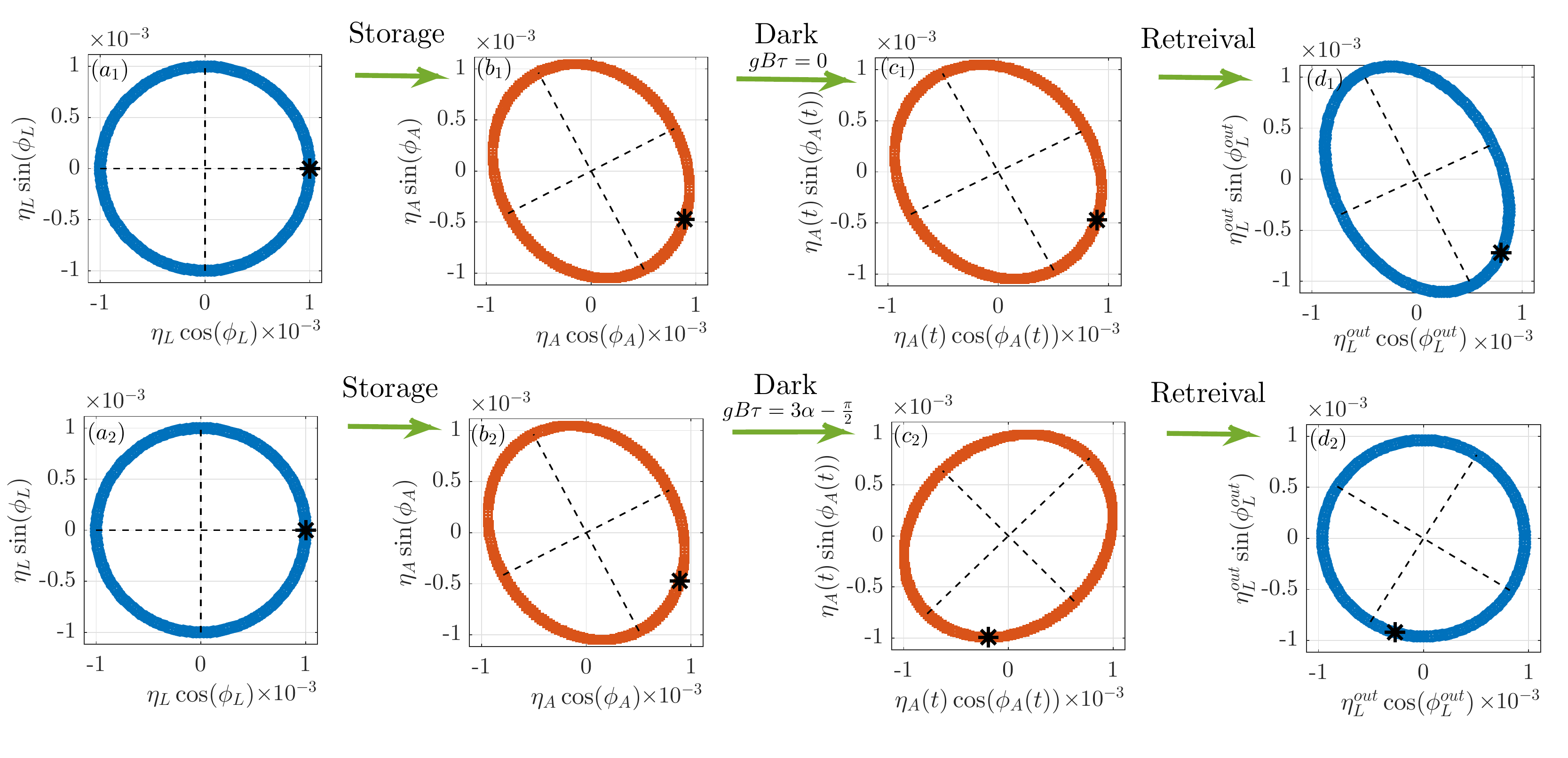}
\par\end{centering}
\caption{{\small{}Full transformation from storage to retrieval ($\eta_{\mathrm{L}}=10^{-3}$
and $\alpha=0.1$). (a1)-(d1): For a vanishing Larmor precession $\omega_{B}t=0$,
the transformation is elliptical. (a2)-(d2): For $\omega_{B}t=3\alpha-\pi/2$,
the elliptical distortion is eliminated. \label{fig: total transformation}}}
\end{figure}

The elliptical distortion can be removed completely by setting the
magnetic field to satisfy $\omega_{B}t\approx3\alpha-\pi/2$, see
Fig.~\ref{fig: total transformation}. For this value, the transformation
simplifies to $\phi_{\mathrm{L}}^{\mathrm{out}}\approx\phi_{\mathrm{L}}-6\alpha$
and $\eta_{\mathrm{L}}=e^{-t/\tau_{\mathrm{s}}}\eta_{\mathrm{L}}$,
such that the amplitude $\eta_{\mathrm{L}}$ is independent of $\phi_{\mathrm{L}}$.
We emphasize that the distortion is eliminated for all orders of $\alpha$
and is thus completely removed. The process can be described in terms
of a perfect ``quantum eraser'', since it neither depends on the
(intermediate) spin states nor it involves any classical feedback
or measurement of the quantum system. 

\begin{thebibliography}{10}
\bibitem{Lukin-RMP-2003}Lukin, M. D. Colloquium: Trapping and manipulating
photon states in atomic ensembles. Rev. Mod. Phys. 75, 457\textendash 472
(2003).

\bibitem{oPTICAL-QUANTUM-COMPUTING-SCIENCE}O'Brien, J. L. Optical
quantum computing. Science 318, 1567\textendash 1570 (2007).

\bibitem{Heshami-2016} Heshami K., Englanda D. G., Humphreysb P.
C., Bustarda P. J., Acostac V. M., Nunn J. \& Sussman B. J. Quantum
memories: emerging applications and recent advances. J. Mod. Opt.
63, 2005\textendash 2028 (2016).

\bibitem{Tittel-review-nature-photonics} Lvovsky, A. I., Sanders,
B. C. \& Tittel, W. Optical quantum memory. Nature Photon. 3, 706\textendash 714
(2009).

\bibitem{Gisin-repeaters} Sangouard, N., Simon, C., De Riedmatten,
H. \& Gisin, N. Quantum repeaters based on atomic ensembles and linear
optics. Rev. Mod. Phys. 83, 33\textendash 80 (2011). 

\bibitem{Firstenberg-FLAME} Finkelstein R., Poem E., Michel O., Lahad
O. \& Firstenberg O. Fast, noise-free memory for photon synchronization
at room temperature, arXiv:1708.01919 (2017).

\bibitem{Polzik-RMP-2010} Hammerer, K., Sørensen, A. S. \& Polzik,
E. S. Quantum interface between light and atomic ensembles. Rev. Mod.
Phys. 82, 1041\textendash 1093 (2010).

\bibitem{Buchler-GEM}Hosseini M., Sparkes B. M., Campbell G., Lam
P. K. \& Buchler B. C. High efficiency coherent optical memory with
warm rubidium vapour. Nat. Commun. 2, 174 (2011).

\bibitem{Buchler-efficiency-storageTime-review} Cho Y.-W., Campbell
G. T., Everett J. L., Bernu J., Higginbottom D. B., Cao M. T., Geng
J., Robins N. P., Lam P. K. \& Buchler B. C. Highly efficient optical
quantum memory with long coherence time in cold atoms, Optica 3, 100-107
(2016).

\bibitem{Novikova-review} Novikova, I., Walsworth, R. \& Xiao, Y.
Electromagnetically induced transparency-based slow and stored light
in warm atoms. Laser Photon. Rev. 6, 333\textendash 353 (2012).

\bibitem{Lukin-PRL-2001} Phillips, D. F., Fleischhauer, A., Mair,
A., Walsworth, R. L. \& Lukin, M. D. Storage of light in atomic vapor.
Phys. Rev. Lett. 86, 783\textendash 786 (2001).

\bibitem{Weitz-Spinor-polaritons} Karpa L., Vewinger F. \& Weitz
M. Resonance Beating of Light Stored Using Atomic Spinor Polaritons.
Phys. Rev. Lett. 101, 170406 (2008).

\bibitem{Novikova-integrated-gain} Novikova I., Phillips D. F. \&
Walsworth R. L. Slow Light with Integrated Gain and Large Pulse Delay.
Phys. Rev. Lett. 99, 173604 (2007).

\bibitem{Halfmann-EIT-solid-state} Schraft D., Hain M., Lorenz N.
\& Halfmann T. Stopped Light at High Storage Efficiency in a $\text{Pr}^{3+}:\text{Y}_{2}\text{SiO}_{5}$
Crystal. Phys. Rev. Lett. 116, 073602 (2016).

\bibitem{Happer-1972} Happer, W. Optical pumping. Rev. Mod. Phys.
44, 169\textendash 249 (1972).

\bibitem{Optimal-light-storage-Novikova-Gorshkov} Phillips, N. B.,
Gorshkov, A. V. \& Novikova, I. Optimal light storage in atomic vapor.
Phys. Rev. A 78, 023801 (2008).

\bibitem{Happre-SERF} Happer, W. \& Tang, H. Spin-exchange shift
and narrowing of magnetic resonance lines in optically pumped alkali
vapors. Phys. Rev. Lett. 31, 273\textendash 276 (1973).

\bibitem{Katz-nonlinear-SERF} Katz, O. Dikopoltsev M., Peleg O.,
Shuker M., Steinhauer J. \& Katz N. Nonlinear elimination of spin-exchange
relaxation of high magnetic moments. Phys. Rev. Lett. 110, 263004
(2013).

\bibitem{Optical-magnetometery-Romalis} Budker, D. \& Romalis, M.
Optical magnetometry. Nature Phys. 3, 227\textendash 234 (2007). 

\bibitem{160-att-T-Romalis} Dang, H. B., Maloof, A. C. \& Romalis,
M. V. Ultrahigh sensitivity magnetic field and magnetization measurements
with an atomic magnetometer. Appl. Phys. Lett. 97, 151110 (2010).

\bibitem{Happer-clock}Jau Y.Y., Post A.B., Kuzma N.N., Braun A.M.,
Romalis M.V. \& Happer W. Intense, narrow atomic-clock resonances.
Phys. Rev. Lett. 92, 110801 (2004). 

\bibitem{optically-pumped-atoms-book} Happer W., Jau Y. Y. \& Walker
T. Optically pumped atoms. 159\textendash 218 (WILEY-VCH Press, 2010).

\bibitem{Polzik-2002} Schori, C., Julsgaard, B., Sørensen, J. L.
\& Polzik, E. S. Recording quantum properties of light in a long-lived
atomic spin state: Towards quantum memory. Phys. Rev. Lett. 89, 057903
(2002).

\bibitem{Goldfarb-CPO} Neveu P., Maynard M. A., Bouchez R., Lugani
J., Ghosh R., Bretenaker F., Goldfarb F. \& Brion E. Coherent Population
Oscillation-Based Light Storage. Phys. Rev. Lett. 118, 073605 (2017).

\bibitem{Budker-Yavchuk-Self-Rotation-2001} Rochester S.M., Hsiung
D.S., Budker D., Chiao R.Y. , Kimball D.F. \& Yashchuk V.V. Self-rotation
of resonant elliptically polarized light in collision-free rubidium
vapor. Phys. Rev. A 63, 043814 (2001).

\bibitem{alkali_curing}Li, W., Balabas, M., Peng X., Pustelny, S.,
Wickenbrock, A., Yang Y., Guo H. \& Budker, D. Investigation of antirelaxation
wall coatings beyond melting temperatures,\textquotedbl{} 2017 Conference
on Lasers and Electro-Optics (CLEO), San Jose, CA, 2017, pp. 1-2.

\bibitem{SEOP-Happer-RMP} Walker, T. G. \& Happer, W. Spin-exchange
optical pumping of noble-gas nuclei. Rev. Mod. Phys. 69, 629\textendash 642
(1997).

\bibitem{Haroce-1970-coherent-coupling} Haroche, S. \& Cohen-Tannoudji,
C. Resonant transfer of coherence in nonzero magnetic field between
atomic levels of different $g$ factors. Phys. Rev. Lett. 24, 974
(1970). 

\bibitem{katz-coherent-coupling} Katz, O., Peleg, O. \& Firstenberg,
O. Coherent coupling of alkali atoms by random collisions. Phys. Rev.
Lett. 115, 113003 (2015).

\bibitem{Kornack-2002} Kornack, T. W. \& Romalis, M. V. Dynamics
of two overlapping spin ensembles interacting by spin exchange. Phys.
Rev. Lett. 89, 253002 (2002).

\bibitem{Firstenberg-residual-doppler} Firstenberg, O., Shuker M.,
Pugatch R., Fredkin D. R., Davidson N. \& Ron A. Theory of thermal
motion in electromagnetically induced transparency: Diffusion, Doppler,
Dicke and Ramsey. Phys. Rev. A 77, 043830 (2008).\clearpage{}\newpage{}
\end{thebibliography}

\begin{thebibliography}{10}
\bibitem{footnote1}Non-Hermitian Hamiltonian dynamics with stochastic
quantum jumps is used to solve open-systems dynamics \cite{Fleischhauer-EIT-RMP-SI}.
In EIT, population of the excited state is avoided, thus the effect
of the quantum jumps in determining the dark state is small. 

\bibitem{Fleischhauer-EIT-RMP-SI} Fleischhauer, M., Imamoglu, A.
\& Marangos, J. P. Electromagnetically induced transparency: Optics
in coherent media. Rev. Mod. Phys. 77, 633\textendash 673 (2005).

\bibitem{Steck-Cs-SI} Steck, D. A. http://steck.us/alkalidata (2009).

\bibitem{Gorshkov-reversible} Novikova, I., Gorshkov A. V., Phillips
D. F., Sørensen A. S., Lukin M. D. and Walsworth R. L., Optimal Control
of Light Pulse Storage and Retrieval, Phys. Rev. Lett., 98, 243602
(2007).

\bibitem{Nunn-reversible} Nunn J., Reim K., Lee K. C., Lorenz V.
O., Sussman B. J., Walmsley I. A. and Jaksch D., Multimode Memories
in Atomic Ensembles, Phys. Rev. Lett. 101, 260502 (2008). 

\bibitem{optically-pumped-atoms-book-SI} Happer W., Jau Y. Y. \&
Walker T. Optically pumped atoms. 159\textendash 218 (WILEY-VCH Press,
2010).

\bibitem{Happer-Tang-suceptability-1970}B. S. Mathur, H. Y. Tang,
and W. Happer, Phys. Rev. A 2, 648 (1970).

\bibitem{footnote2}When identifying the atomic quasi-dark state in
the analysis of the storage process, we assumed time-invariant light
fields $\Omega_{\mathrm{c}}$ and $\Omega_{\mathrm{s}}$. To identify
the least decaying mode of the light, we correspondingly take a time-invariant
(steady-state) atomic susceptibility $\left\langle \overleftrightarrow{\chi}\right\rangle $.

\bibitem{novikova-gain-SI} Novikova, I., Phillips, D. F. \& Walsworth.,
R. L. Slow light with integrated gain and large pulse delay Phys.
Rev. Lett. 99, 173604 (2007).

\bibitem{Xiao-phase-EIT-SI}Xu, X., Shen, S. \& Xiao, Y. Tuning the
phase sensitivity of a double-lambda system with a static magnetic
field. Opt. Exp. 21, 11705 (2013).
\end{thebibliography}
\end{document}